\documentstyle[12pt]{article}
\begin{document}
\title{AdS/CFT Correspondence and \\ 
Quotient Space Geometry}
\author{Yi-hong Gao\thanks{e-mail: 
{\tt gaoyh@itp.ac.cn}} \vspace{.3cm}
\\ {\it Institute of Theoretical Physics} \\ 
{\it Beijing 100080, China}}
\date{}
\maketitle
\baselineskip .63cm
\vskip 1.2cm
\begin{center} 
{\bf Abstract} 
\vskip .8cm
\begin{minipage}{152mm}
\baselineskip .63cm
We consider a version of the $AdS_{d+1}/CFT_{d}$ correspondence, in 
which the bulk space is taken to be the quotient manifold $AdS_{d+1}
/\Gamma$ with a fairly generic discrete group $\Gamma$ acting 
isometrically on $AdS_{d+1}$. We address some geometrical issues 
concerning the holographic principle and the UV/IR relations. It 
is shown that certain singular structures on the quotient boundary 
${\bf S}^{d}/\Gamma$ can affect the underlying physical spectrum. 
In particular, the conformal dimension of the most relevant operators 
in the boundary CFT can increase as $\Gamma$ becomes ``large''. 
This phenomenon also has a natural explanation in terms of the bulk 
supergravity theory. The scalar two-point function is computed using 
this quotient version of the AdS/CFT correspondence, which agrees 
with the expected result derived from conformal invariance 
of the boundary theory.
\end{minipage}
\end{center}
\newpage
\voffset -.6in
\renewcommand{\theequation}{\thesection.\arabic{equation}}
\csname @addtoreset\endcsname{equation}{section}
\section{Introduction}
The AdS/CFT correspondence proposed in \cite{Ma1}\cite{GKP}\cite{Wi1}
gives a surprisingly powerful method to investigate strong coupling 
physics. In its simplest form, this proposal relates tree-level 
supergravity on $(d+1)$-dimensional anti-de Sitter space to a 
conformal field theory on the round sphere ${\bf S}^{d}$, which is 
the conformal boundary of $AdS_{d+1}$. One may consider CFT on a more 
complicated manifold $M$ as well, by replacing the bulk spacetime 
$AdS_{d+1}$ with a negatively curved Einstein space $X$ of boundary 
$M$. In fact, using topologically non-trivial $X$ can provide 
additional tests for the bulk/boundary correspondence 
\cite{Wi2}\cite{AW}\cite{Wi3}.

There are several ways of constructing a non-trivial Einstein 
manifold with negative cosmological constant. One way is to place 
black holes in AdS space. This amounts to a study of the boundary 
theory at finite temperature \cite{Wi2}. While this construction 
can exhibit interesting phase structures in the large $N$ limit 
(using a mechanism discovered by Hawking and Page \cite{HP} many 
years ago), finding black hole solutions in the AdS background 
is however by no means an easy task; at present only a few such 
solutions are explicitly known. 

Alternatively, we can pick a discrete subgroup $\Gamma$ of $SO(d,2)$ 
that acts isometrically on $AdS_{d+1}$, and take the quotient 
manifold $X=AdS_{d+1}/\Gamma$ to be the bulk spacetime. The 
resulting space will constitute a $(d+1)$-dimensional negatively 
curved Einstein manifold, on which the classical supergravity action 
can be defined. One thus expects that gravity in $X$ still
corresponds to some conformal field theory on the quotient boundary 
$M={\bf S}^{d}/\Gamma$. For a check on this quotient version of the 
AdS/CFT correspondence, recall that the Hawking-Page reference space 
(denoted by $X_{1}$ in \cite{Wi1}\cite{Wi2}) takes the form $AdS_{d+
1}/{\bf Z}$, which is associated to a boundary CFT in low-temperature 
phase. Another useful check concerns a quotient construction of the 
BTZ black hole \cite{BTZ}\cite{btzh} in $AdS_{3}$, where $\Gamma$ is 
generated by a single loxodromic element; supergravity in this 
geometry is again related to a $2D$ CFT (see e.g. \cite{Mar} for such 
a study and the references therein). Note that these examples made a 
common choice of $\Gamma$ in elementary discrete groups. 

In the present work we wish to consider some general features of the 
quotient AdS/CFT correspondence. We will address a couple of 
geometrical issues that are closely related to the holographic 
principle \cite{holo} and the UV/IR relations \cite{SW}\cite{PP} 
between bulk- and boundary- theories. As we will see in detail, 
certain singular structures on the quotient boundary ${\bf S}^{
d}/\Gamma$ may affect the underlying physical spectrum. This effect 
could not be seen by choosing $\Gamma$ in elementary discrete groups; 
mathematically such a choice is too special to describe generic 
structures of quotient manifolds. Thus, in this paper we will take 
$\Gamma$ to be non-elementary, namely it contains no abelian 
subgroups of finite index. With this choice the conformal dimension 
of the most relevant operators in the boundary CFT will depend 
nontrivially on $\Gamma$, provided $\Gamma$ is ``large'' enough. 
We will give some mathematical explanations of such a dependence, 
both on the supergravity side and from the boundary point of view. 
According to this dependence we could have, for $\Gamma$ extremely 
large, a boundary theory that does not contain any relevant operators, 
--- corresponding to a bulk supergravity theory without tachyon modes.

The existence of singular structures on the boundary and its relations 
to the bulk geometry have been extensively studied by mathematicians. 
Some of the mathematical results will be reviewed in this paper. 
Section 2 provides a brief description of the quotient space geometry. 
In section 3, we discuss several implications of the holographic 
principle as well as certain geometrical subtleties arising from 
constructing the quotient boundary. Section 4 then deals with scalar 
conformal fields on the quotient boundary, where we derive a constraint 
on the spectrum of conformal dimensions. In section 5 we compute the 
scalar two-point function using the quotient AdS/CFT correspondence. 
Finally, after presenting our conclusions in section 6, we give an
appendix to clarify some points in the text.
\section{Taking the Quotient}
We will acturally work in Euclidean language and 
take the covering space of our bulk to be the $(d+1)$-dimensional 
hyperbolic space ${\bf H}^{d+1}\cong SO(d+1,1)/SO(d+1)$. This is a 
complete, simply connected Riemannian manifold with negative 
constant curvature and having $SO(d+1,1)$ as its isometry group. 
A particularly useful description of ${\bf H}^{d+1}$ is the open 
half-space model, which has the standard hyperbolic metric $ds^{2
}=|dx|^{2}/x_{0}^{2}$ and the boundary sphere 
\begin{equation}
{\bf S}^{d}\;\simeq\;{\bf R}^{d}\cup\{\infty\}\;\equiv\; 
\hat{{\bf R}}^{d}
\label{ebd}
\end{equation}
at infinity, here the $d$-dimensional plane ${\bf R}^{d}$ is 
specified by the equation $x_{0}=0$. 

Geometrically, there is a simple realization of isometric 
transformations in this half-space model \cite{bear1}. 
Suppose that $\cal S$ is a $d$-dimensional Euclidean 
hemisphere in the half-space whose equatorial plane coincides 
with the boundary plane ${\bf R}^{d}$ in (\ref{ebd}). The 
inversion in $\cal S$, which will be denoted by 
${\cal I}_{\cal S}$, is an orientation-reversing isometry 
of ${\bf H}^{d+1}$. In the limiting case when $\cal S$ 
approaches a $d$-dimensional Euclidean half-plane in 
${\bf H}^{d+1}$ perpendicular to the boundary plane, the 
inversion ${\cal I}_{\cal S}$ becomes an ordinary reflection, 
which again is an orientation-reversing isometry. To obtain an 
orientation-preserving isometry, one simply takes product 
of an even number of such ${\cal I}_{\cal S}$'s. It 
turns out that each isometry of the hyperbolic space can be 
realized in this way, namely for any $g\in SO(d+1,1)$, we can 
find $2n$ hemispheres 
${\cal S}_{1}, {\cal S}_{2}, \cdots, {\cal S}_{2n}$ 
as described above (possibly including those in the Euclidean 
half-plane limit), so that
\begin{equation}
g\;=\;{\cal I}_{{\cal S}_{1}}{\cal I}_{{\cal S}_{2}}
\cdots{\cal I}_{{\cal S}_{2n}}.
\label{product}
\end{equation}
Note that each inversion 
${\cal I}_{{\cal S}}:{\bf H}^{d+1}
\rightarrow{\bf H}^{d+1}$ on the hyperbolic space induces a 
transformation ${\cal I}_{\partial\bar{\cal S}}:{\bf S}^{d}
\rightarrow{\bf S}^{d}$ on the boundary (\ref{ebd}), defined 
by the inversion in $\partial\bar{\cal S}$, where $\bar{\cal S}$
is the closure of $\cal S$ whose boundary $\partial\bar{\cal S}$ 
is a $(d-1)$-dimensional Euclidean sphere (or Euclidean plane in 
the limiting case) living in (\ref{ebd}). Thus group elements of 
the form (\ref{product}) also act naturally on the boundary sphere, 
giving the so-called M\"{o}bius transformations \cite{bear1}. 

Now we choose a discrete subgroup $\Gamma$ of $SO(d+1,1)$. 
Discreteness means that each $\gamma\in\Gamma$ has a neighborhood 
$U_{\gamma}\subset SO(d+1,1)$ such that $U_{\gamma}\cap\Gamma=
\{\gamma\}$. So every sequence of infinitely many distinct elements 
$\{\gamma_{n}\}$ in $\Gamma$ cannot converge to any point of 
$SO(d+1,1)$ and, in particular, we have $\lim\limits_{n\rightarrow
\infty}||\gamma_{n}||=\infty$. Evidently, any element of $\Gamma$ 
has a product represention (\ref{product}) in terms of inversions. 
This specifies how $\Gamma$ acts on ${\bf H}^{d+1}$ in the half-space 
description. The $(d+1)$-dimensional bulk spacetime $X$ is then 
constructed by
\begin{equation}
X\;=\;{\bf H}^{d+1}/\Gamma.
\label{bulk}
\end{equation}
If $\Gamma$ does not contain elements of finite order, then it 
acts freely on ${\bf H}^{d+1}$, so that $X$ is a smooth hyperbolic 
manifold without orbifold singularities.

Since $\Gamma$ acts isometrically on ${\bf H}^{d+1}$, the local 
geometry of $X$ looks the same as the original hyperbolic geometry. 
In particular $X$ inherits the standard hyperbolic metric from 
${\bf H}^{d+1}$. Of course, taking the quotient (\ref{bulk}) will 
generally break the isometry group $SO(d+1,1)$. The residual 
spacetime symmetry on $X$ will be determind 
by\footnote{A quick derivation of this group is as follow. Each 
point $x$ of $X$ can be considered as an orbit $[y]\equiv\Gamma
\cdot y$ in ${\bf H}^{d+1}$, starting at some $y$ in the hyperbolic 
space. The action of $g\in SO(d+1,1)$ on $X$ is defined by 
$x \mapsto g\cdot x\equiv [g\cdot y]$, which is well-defined 
provided $[g\cdot y]$ is independent of the choice of representive 
$y$ in the orbit; thus, if $y'=\gamma\cdot y$ is another 
representive, there should exist a $\gamma'\in\Gamma$ such that 
$g\cdot y'=\gamma'\cdot(g\cdot y)$. The collection of all such 
$g$ is precisely the isometry group of $X$.}:
\begin{equation}
G\;=\;\{g\in SO(d+1,1)\;\;\;|\;\;g\cdot\gamma\cdot g^{-1}
\;\in\;\Gamma,\;\;\forall\;\gamma\;\in\;\Gamma\}.
\label{C-symm}
\end{equation}
In the quotient version of the AdS/CFT correspondence, this 
isometry group should be identified with the unbroken conformal 
symmetry of the boundary theory. An independent verification of 
this fact will be given at the end of section 5. 

From the boundary point of view, it is natural to regard $\Gamma$ 
as a discrete group of certain M\"{o}bius transformations acting 
on the boundary sphere ${\bf S}^{d}$. Thus the ``quotient boundary''
is constructed by
\begin{equation}
M\;=\;{\bf S}^{d}/\Gamma.
\label{boundary}
\end{equation}
In contrast to the previous case, $M$ in general does not inherit 
any metric from ${\bf S}^{d}$, as generic M\"{o}bius transformations 
cannot preserve such a metric. Nevertheless, this quotient space 
{\it does} inherit a conformal structure from ${\bf S}^{d}$.
\section{Holography and Singular Structures}
The usual AdS/CFT correspondence associates each (on-shell) bulk degree 
of freedom on ${\bf H}^{d+1}$ to a certain boundary degree of freedom 
on ${\bf S}^{d}$. When taking the quotient by $\Gamma$, bulk fields in
$X={\bf H}^{d+1}/\Gamma$ can be obtained by $\Gamma$-invariant 
projection of those in the original AdS space, and the resulting theory 
may be considered as a truncated version of the ordinary AdS supergravity, 
in which only $\Gamma$-invariant degrees of freedom are allowed. 
The projection also forces boundary values of a bulk field to be 
invariant under $\Gamma$, which, in turn, give rise to a 
$\Gamma$-invariant dual image in the boundary CFT defined on the original
sphere ${\bf S}^{d}$, via the usual AdS/CFT duality. The collection 
of all such dual images then forms the field content of a truncated 
boundary theory. A naive generalization of the AdS/CFT correspondence
will thus provide a connection between the truncated bulk- and 
boundary- theories. Although this could not be checked directly by taking
the near-horizon limit (a generic $\Gamma$ does not preserve the metrics 
generated by D brane sources, so taking the quotient in general cannot 
commute with taking the near-horizon limit), explicit computations
of correlators support such a generalization; see appendix A.1 for a 
detailed discussion.

At this point, one might expect that supergravity on the quotient bulk 
$X$ should correspond to a conformal field theory living in the 
$d$-dimensional space $M$, defined by (\ref{boundary}), since the latter 
theory (if exists) seems to provide the natural $\Gamma$-invariant truncation 
of the original boundary theory on ${\bf S}^{d}$. The geometry of the 
quotient boundary space (\ref{boundary}) is quite clear in examples with 
elementary discrete groups \cite{Wi1}\cite{Wi2}\cite{Mar}. For generic 
$\Gamma$, however, there may exist some subtleties that at first sight
will raise a puzzle, --- ignoring such subtleties and simply supposing 
$M$ to be the space on which the holographic boundary theory lives would
violate the UV/IR relations \cite{Wi1}\cite{SW}\cite{PP} between 
boundary- and bulk- theories and thus conflict with the holographic 
principle proposed in \cite{holo}. We will see how this can happen in 
the following extreme case. 

Generally speaking, taking the quotient of ${\bf H}^{d+1}$ by a 
discrete group $\Gamma$ will cause reduction of the spacetime 
volume $V\rightarrow V/|\Gamma|$. For finite groups this reduction 
is invisible because of $V=\infty$ in the hyperbolic metric. But 
when $\Gamma$ is chosen to be large enough, we can actually make 
the quotient space volume $Vol(X)$ finite. In this case $X$ may be 
viewed as a regulated version of the original AdS-space 
${\bf H}^{d+1}$, with a certain $\Gamma$-invariant infrared cutoff 
imposed. In other words, one regulates the bulk theory by putting 
it into some finite-volume box --- a fundamental domain for $\Gamma$, 
and imposing the ``periodic'' conditions compatible with the action
of $\Gamma$ on each side of that box. The infrared cutoff may be 
thought of as the mean length scale of the fundamental domain, thus 
having a $\Gamma$-invariant meaning. 
On the other hand, we may start from a classical conformal field theory 
defined on the original AdS-boundary ${\bf S}^{d}$. The quantum version 
of this theory can be regulated by an ultraviolet cutoff so that it 
lives on a discrete subspace of ${\bf S}^{d}$. There may exist many 
different lattice models of the boundary sphere, but according to the 
UV/IR relations \cite{Wi1}\cite{SW}\cite{PP}, only one such discrete 
subspace $M'$ can correspond to the given infrared regulating 
${\bf H}^{d+1}\leadsto X$ of the bulk, as long as $Vol(X)<\infty$
\footnote{If $Vol(X)=\infty$, $X$ does not really regulate 
${\bf H}^{d+1}$ through an infrared cutoff, so $M'$ needs not to be 
discrete.}. Since this $M'$ arises from an ultraviolet cutoff of 
${\bf S}^{d}$ while $M={\bf S}^{d}/\Gamma$ does not, in general we should 
have $M'\neq M$ and, since the cutoff has a $\Gamma$-invariant meaning, 
$M'$ is likely to be a subset of $M$. Thus, supergravity on the quotient 
bulk space $X$ should generally correspond to a ``boundary'' theory on 
$M'$ other than the quotient boundary (\ref{boundary}).

There is an interesting consequence of the above discussion. 
If $Vol(X)<\infty$, the ultraviolet cutoff completely breaks 
the conformal algebra on ${\bf S}^{d}$, so holography predicts 
that the residual symmetry (\ref{C-symm}) will at most constitute 
a discrete group. This is in accordance with the mathematical 
speculation that generic hyperbolic manifolds with finite volumes 
admit no continuous isometries. Continuous isometries can arise 
only if the volume of $X$ is infinite. 

So far we have only considered an extreme case that can occur 
in the AdS/CFT correspondence: When $Vol(X)<\infty$, one naively 
expects that conformal fields should be defined on the quotient 
boundary $M={\bf S}^{d}/\Gamma$, but turning to the quantum theory 
a rather intriguing, apparantly discrete space $M'\subset M$ can 
possibly emerge as the regulated spacetime. It would be of 
interest to have a geometrical understanding of why the boundary 
of $X$ can be discretized as one deforms $\Gamma$ from generic 
position into this extreme case. 

The crucial point is that the action of a discrete $\Gamma\subset 
SO(d+1,1)$ on ${\bf H}^{d+1}$ may behave very differently from 
that on the boundary sphere ${\bf S}^{d}$. One can show that 
$\Gamma$ always acts discontinuously\footnote{We say that $\Gamma$ 
acts discontinuously on a space $Y$ provided for every compact 
subset $K$ of $Y$, the condition $\gamma(K)\cap K=\emptyset$ 
holds except for finitely many $\gamma\in\Gamma$. Thus a finite 
group acts on $Y$ discontinuously. If $\Gamma$ is infinite and acts 
discontinuously on $Y$, then for all $y\in Y$ as well as $\gamma_{1},
\gamma_{2},\dots\in\Gamma$, the sequence $\{\gamma_{n}(y)\}$ cannot 
converge to any point of $Y$. See section 5.3 of \cite{bear1} for 
a more detailed discussion.} on ${\bf H}^{d+1}$. In fact, if this 
is not the case, then we can find certain $x\in {\bf H}^{d+1}$ and 
a sequence of infinitely many distinct elements $\{\gamma_{n}\}$ 
in $\Gamma$, such that $\gamma_{n}(x)$ coverges to some 
$y\in{\bf H}^{d+1}$. Since ${\bf H}^{d+1}$ can be regarded as the 
coset space $SO(d+1,1)/SO(d+1)$, the points $x$ and $y$ in the 
hyperbolic space may be represented by some group elements 
$g_{x}\sim g_{x}\cdot h_{x}$ and $g_{y}\sim g_{y}\cdot h_{y}$ 
in $SO(d+1,1)$, respectively, up to arbitrary $h_{x},h_{y}\in 
SO(d+1)$. Thus, the limit $\gamma_{n}(x)\rightarrow y$ can be 
alternatively described by $\gamma_{n}\cdot 
g_{x}=g_{\gamma_{n}(x)}\cdot h_{n}$ with 
$g_{\gamma_{n}(x)}\rightarrow g_{y}$ and $h_{n}\in SO(d+1)$. This 
clearly gives $\gamma_{n}=g_{\gamma_{n}(x)}\cdot h_{n}
\cdot g_{x}^{-1}$. Now because $g_{\gamma_{n}(x)}$ has a finite 
limit $g_{y}$ as $n\rightarrow\infty$, and because the $h_{n}$'s 
belong to a compact group $SO(d+1)$, we see that $||\gamma_{n}||$ 
is bounded from above. But this is impossible due to discreteness 
of $\Gamma$. So as far as $\Gamma$ is discrete, its action on 
${\bf H}^{d+1}$ is necessarily discontinuous\footnote{See 
\cite{bear1} for a somewhat different proof of this result in 
the case $d=2$.}. Accordingly the quotient space $X={\bf H}^{d+1}
/\Gamma$ with its natural quotient topology is always a Hausdorff 
space.

On the contrary, a discrete group $\Gamma$ in general does not act 
discontinuously on the boundary of ${\bf H}^{d+1}$ and there may 
exist accumulation (or cluster) points in each $\Gamma$-orbit, 
namely, given any $x$ on the boundary sphere ${\bf S}^{d}$, one 
may find an infinite sequence $\{\gamma_{n}\}$ in $\Gamma$ such 
that $\lim\limits_{n\rightarrow\infty}\gamma_{n}(x)\in{\bf S}^{d}$. 
These points may render the quotient space ${\bf S}^{d}/\Gamma$ 
possible to be non-Hausdorff. The set of all such accumulation 
points is called the limit set $\Lambda_{\Gamma}$ of $\Gamma$. 
This set is a closed, $\Gamma$-invariant subset of ${\bf S}^{d}$; 
actually it is the smallest non-empty subset having such properties. 
Thus $\Lambda_{\Gamma}$ is contained in the closure of an orbit 
${\Gamma\cdot x}$, with $x$ either on the boundary sphere or in 
the interior of the hyperbolic space. Note that if $\Gamma'\subset
\Gamma$, then $\Lambda_{\Gamma'}\subset\Lambda_{\Gamma}$. Hence 
a ``large'' $\Gamma$ actually means that $\Lambda_{\Gamma}$ has 
a relatively large Hausdorff dimension. 

Let us give an example to illustrate how the limit set of $\Gamma$ 
can have positive Hausdorff dimension. This example comes from a 
slight modification of the material exposed in \cite{bear2}. As 
we have mentioned before, the action of $\Gamma$ on the hyperbolic 
space is generated by inversions with respect to some $d$-dimensional 
Euclidean hemispheres (or half-planes) in ${\bf H}^{d+1}$. We 
restrict ourselves to a simple case in which only finitely many such 
hemispheres can occur. Thus, for $i=1,2,\dots,K+1$, let ${\cal S}_{i}$ 
be a $d$-dimensional Euclidean hemisphere with center $a_{i}$ and 
Euclidean radius $r_{i}$, and let ${\cal B}_{i}$ be the 
($d+1$)-dimensional Euclidean half-ball bounded by ${\cal S}_{i}$, 
whose closure is denoted by $\bar{\cal B}_{i}$. Suppose that these 
half-balls are mutually disjoint, namely for each pair $(i,j)$, 
$|a_{i}-a_{j}| > r_{i}+r_{j}$. Then the discrete group $\Gamma$ 
which we will consider consists of all elements of the form 
\begin{equation}
\gamma_{i_{1}i_{2}\dots i_{2n}}\;=\;{\cal I}_{i_{1}}
{\cal I}_{i_{2}}\cdots {\cal I}_{i_{2n-1}}{\cal I}_{i_{2n}},
\;\;\;\;\;\;(i_{1},i_{2},\dots,i_{2n})\in\Sigma(2n)
\label{prod}
\end{equation}
where ${\cal I}_{i}$ is the inversion in ${\cal S}_{i}$, 
$\Sigma(m)$ denotes the set of all sequences $i_{1},\dots,i_{m}$ 
such that $1\leq i_{1},i_{2},\dots,i_{m}\leq K+1$ and $i_{1}\neq 
i_{2}, i_{2}\neq i_{3},\dots, i_{m-1}\neq i_{m}$ (that is, 
adjacent indices have different values). Since all the half-balls 
do not touch each other, we have  ${\cal I}_{i}(\bar{\cal B}_{j})
\subset \bar{\cal B}_{i}$ for $i\neq j$, and therefore 
\begin{equation}
\bar{\cal B}_{i_{1}i_{2}\dots i_{2n+1}}\;\equiv\;
\gamma_{i_{1}i_{2}\dots i_{2n}}(\bar{\cal B}_{i_{2n+1}})
\;\subset\;\bar{\cal B}_{i_{1}},\;\;\;\;\;\;\forall\;\;
(i_{1},i_{2}\dots i_{2n+1})\;\in\;\Sigma(2n+1).
\label{sball}
\end{equation}
Moreover, it is not difficult to see that (\ref{sball}) itself 
is a closed half-ball in ${\bf H}^{d+1}\cup {\bf S}^{d}$, with 
the center $a_{i_{1}i_{2}\dots i_{2n+1}}$ and radius 
$r_{i_{1}i_{2}\dots i_{2n+1}}$ determined iteratively by
\begin{equation}
a_{i_{1}i_{2}\dots i_{m}}\;=\;a_{i_{1}}-\frac{r_{i_{1}}^{2}
\;(a_{i_{1}}-a_{i_{2}\dots i_{m}})}{|a_{i_{1}}-a_{i_{2}
\dots i_{m}}|^{2}-r_{i_{2}\dots i_{m}}^{2}},\;\;\;\;\;\;
r_{i_{1}i_{2}\dots i_{m}}\;=\;\frac{r_{i_{1}}^{2}\;r_{i_{2}
\dots i_{m}}}{|a_{i_{1}}-a_{i_{2}\dots i_{m}}|^{2}-r_{i_{2}
\dots i_{m}}^{2}}
\label{a&r}
\end{equation}
where $m\geq 2$ and $(i_{1},i_{2},\dots,i_{m})\in\Sigma(m)$. 
It follows from the (geometrically obvious) inequalities 
$|a_{i_{1}}-a_{i_{2}\dots i_{m}}|^{2}-r_{i_{2}\dots i_{m}}^{2}
\geq (|a_{i_{1}}-a_{i_{2}\dots i_{m}}|-r_{i_{2}\dots i_{m}})^{2}\geq
(|a_{i_{1}}-a_{i_{2}}|-r_{i_{2}})^{2}$ that the radius of $\bar{\cal 
B}_{i_{1}i_{2}\dots i_{2n+1}}$ will shrink to zero when $n$ tends to 
infinity. Thus, the following closed set
\begin{equation}
E\;=\;\bigcap\limits_{n=1}^{\infty}\;\;\bigcup\limits_{(i_{1}i_{2}
\dots i_{2n+1})\;\in\;\Sigma(2n+1)}\;\;\bar{\cal B}_{i_{1}i_{2}
\dots i_{2n+1}}
\label{limit}
\end{equation}
is a subset of the boundary sphere ${\bf S}^{d}$ and, by construction, 
it is the smallest non-empty closed subset of ${\bf S}^{d}$ invariant 
under the action of $\Gamma$. So $E$ is precisely the limit set 
$\Lambda_{\Gamma}$. Since (\ref{limit}) constitutes a kind of 
the ``general Cantor sets'' \cite{bear3}, its Hausdorff dimension 
$d_{H}(E)$ obeys \cite{bear2}
\begin{equation}
d_{H}(E)\;\geq C_{K}\log K 
\end{equation}
where $C_{K}$ is some positive constant. Thus, for $K\geq 2$, the 
limit set of $\Gamma$ indeed has a positive Hausdorff dimension. 
(In the case $K=1$ when $\Gamma$ becomes an elementary group, we 
have $d_{H}(E)=0$ and $E$ consists only of two points.) 

Returning to generic case, we see that $M={\bf S}^{d}/\Gamma$ 
is not a Hausdorff space. It is not quite clear how to quantize 
classical conformal fields on a non-Hausdorff space. So when we 
try to associate a quantized ``boundary theory'' to (semi-) 
classical gravity on the bulk spacetime $X$, we have to discard 
the singular points on $M$ that cause the quotient boundary to 
be non-Hausdorff, and consider the maximal subset $M'$ of $M$ such 
that $M'$ is topologically a Haustorff space. This naturally leads 
us to the notation of {\it domain of discontinuity}, $D_{\Gamma}$, 
defined as the $\Gamma$-invariant open set $D_{\Gamma}={\bf S}^{d} 
- \Lambda_{\Gamma}$. The group $\Gamma$ now acts properly 
disontinuously on $D_{\Gamma}$ and we propose that the ``regulated
boundary'' $M'$ is to be constructed as
\begin{equation}
M'\;=\;D_{\Gamma}/\Gamma,
\label{tbd}
\end{equation}
plus possible cusps of $\Gamma$ living outside $D_{\Gamma}$.
We will call such $M'$ as the ``Kleinian boundary'' of $X$.
Geometrically, the quotient (\ref{tbd}) can be interpreted 
as the boundary of the Kleinian manifold $({\bf H}^{d+1}\cup 
D_{\Gamma})/\Gamma$ \cite{gt3m}. Any point on ${\bf S}^{d}$ living 
in the limit set will come from an interior point $x$ of 
${\bf H}^{d+1}$ under a certain limit $\lim_{n\rightarrow\infty}
\gamma_{n}(x)$, $\gamma_{n}\in\Gamma$. So when taking the quotient 
(\ref{bulk}) by $\Gamma$, a point $y\in\Lambda_{\Gamma}
\subset{\bf S}^{d}$ can hardly be thought of as a boundary point 
of the quotient space $X$, unless $y$ is a cusp. 

Thus, if $\Gamma$ contains no parabolic elements or, more 
generally, if $\Gamma$ has a fundamental domain in ${\bf H}^{d+1}$ 
that does not touch the limit set (so that all possible cusps are 
in $D_{\Gamma}$), then the Kleinian boundary of $X$ is just the
$d$-dimensional Hausdorff space (\ref{tbd}). On the other hand, 
if $\Gamma$ has cusps not living in $D_{\Gamma}$, they should be 
regarded as additional points of the Kleinian boundary $M'$. These 
additional points may become important when $\Lambda_{\Gamma}$ is 
so large that all the connected components of $D_{\Gamma}$ collapse. 
In that case the Hausdorff dimension of the limit set reaches its 
maximal value $d$ and $M'$ is discretized or even becomes empty. 
Typically this can occur when $X$ has finite volume or is compact. 
We thus find a geometrical resolusion of the apparent puzzle 
mentioned at the begining of this section. The puzzle can be 
removed simply by requiring that the quotient version of the AdS/CFT 
correspondence should relate supergravity in $X$ to a holographic 
theory defined on the Kleinian boundary $M'$, and not on the naive 
quotient space $M={\bf S}^{d}/\Gamma$. 

We now elucidate the above discussions by some examples. The most 
familiar example concerns a finitely generated Fuchsian group of 
the first kind \cite{Siegel}. This is a discrete subgroup of 
$PSL(2,{\bf R})$ acting on ${\bf H}^{2}$ as well as on its boundary 
$\hat{\bf R}$, whose limit set $\Lambda_{\Gamma}$ turns out to be 
the whole $\hat{\bf R}$ and hence the domain of discontinuity 
$D_{\Gamma}$ is empty. To get an intuitive picture of what the 
``Kleinian boundary'' of the Riemann surface $X={\bf H}^{2}/
\Gamma$ should look like, let us consider the typical case where the 
canonical fundamental domain $\cal F$ for $\Gamma$ has a finite 
hyperbolic volume. Then according to \cite{Siegel}, $\cal F$ has 
$4g+s+p$ (finitely many) vertices, $4g$ of which are used to create 
genus of $X$, $s$ of which correspond to elliptic fixed points, and 
$p$ of which sit on $\hat{\bf R}$, corresponding to cusps. Each 
elliptic fixed point is associated to a finite cyclic subgroup of 
$\Gamma$, hence giving rise to an orbifold singularity in the 
interior of the Riemann surface. If we pick local coordiates $z$ 
(with $|z| < 1$) at an elliptic fixed point, then a locally analytic 
function $f$ will behave as $f(z)\sim z^{1/\nu}$ at that point, 
where $\nu$ is the order of the cyclic subgroup associated to the 
elliptic fixed point. On the other hand, if we pick local coordinates 
at a cusp, we will find $f(z)\sim\log z$ near that cusp; thus the 
Riemann surface $X$ has $p$ infinitely long tubes and the cusps can 
be thought of as the ends of these tubes at infinity. Intuitively,
the ``boundary'' of $X$ should be identified with the set of cusps, 
in agreement with our construction of the Kleinian boundary since
in this example $D_{\Gamma}=\emptyset$. This discretization was 
predicted earlier by the holographic argument. 

As another example, we consider some torsion-free discrete 
subgroups of $PSL(2,{\bf C})$ acting on ${\bf H}^{3}$ and on 
the boundary sphere ${\bf S}^{2}\cong\hat{\bf C}\equiv {\bf C}
\cup\{\infty\}$. Let us begin with a $\Gamma$ for which the 
limit set $\Lambda_{\Gamma}$ constitutes a circle ${\bf S}^{1}$ 
in the extended complex plane $\hat{\bf C}$ \cite{gt3m}. In this 
case, the domain of discontinuity $D_{\Gamma}=D_{\Gamma}^{+}
\cup D_{\Gamma}^{-}$ has two connected components (disks) 
$D_{\Gamma}^{\pm}$, separated from each other by that circle. 
Since $\Gamma$ acts properly discontinuously on $D_{\Gamma}$, 
the quotient space $M'=D_{\Gamma}/\Gamma$ is a surface with two 
components $M'^{\pm}=D_{\Gamma}^{\pm}/\Gamma$, which inherits a 
comformal structure from $D_{\Gamma}$. Intuitively, this surface 
gives the main body of the ``boundary'' of $X={\bf H}^{3}/\Gamma$ 
and in fact it is the continuous part of the Kleinian boundary.
The volume of $X$ is infinite since $M'$ has nonzero area.

Now we will focus on what can happen when $\Gamma$ is 
deformed into a limiting group $\Gamma^{*}$ so that 
$Vol({\bf H}^{3}/\Gamma^{*})<\infty$. Suppose that $\Gamma^{*}$ 
arises as the fundamental group of a certain Riemann surface 
$M^{*}$. According to \cite{Thur}, under a quasi-conformal 
deformation of $\Gamma$ the limit set $\Lambda_{\Gamma}$ is 
also deformed, from the circle ${\bf S}^{1}$ to a Jordan curve 
in $\hat{\bf C}$, and the conformal structures $j^{\pm}$ of 
$M'^{\pm}$ departure from their original positions in 
Teichmuller space. The limiting group $\Gamma^{*}$ can be 
obtained by moving $j^{\pm}$ simultaneously towards two 
boundary points $j^{\pm}(\infty)$ of Teichmuller space. One 
of the boundary points, say $j^{+}(\infty)$, is determined by 
pinching a lamination $\lambda^{+}$ of $M^{*}$ and, if we pick 
another lamination $\lambda^{-}$ such that 
$\lambda^{\pm}$ fill up $M^{*}$, then by pinching $\lambda^{-}$ 
we can get the other boundary point, $j^{-}(\infty)$. As $j^{\pm}$ 
approach $j^{\pm}(\infty)$, the Jordan curve $\Lambda_{\Gamma}$ 
becomes more and more complicated, --- it can eventually fill the 
whole complex plane when the double limit is reached \cite{Thur}. 
The hyperbolic structure of the final quotient space $X^{*}=
{\bf H}^{3}/\Gamma^{*}$ is typically described by Thurston's 
mapping torus constructed from identifying $(M^{*},1)$ to 
$(\phi(M^{*}),0)$ in the cylinder $M^{*}\times [0,1]$, where 
$\phi:M^{*}\rightarrow M^{*}$ is a diffeomorphism and when it is 
regarded as a transformation of Teichmuller space, it fixes 
precisely the two boundary points $j^{\pm}(\infty)$. Obviously, 
the resulting mapping torus has no boundary if $M^{*}$ is compact. 
This can be expected since $D_{\Gamma^{*}}=\emptyset$ and 
$\Gamma^{*}$ has no cusps for a compact $M^{*}$. Cusps of 
$\Gamma^{*}$ can arise if $M^{*}$ is not compact. The set 
of all such cusps should now be identified with the Kleinian 
boundary of $X^{*}$, and thus we have a concrete model to see 
how $M'$ can be discretized during the limiting process 
$\Gamma\rightarrow\Gamma^{*}$. Again, these geometrical 
considerations are consistent with the holographic prediction 
given at the begining of this section: finite-volume bulk space 
has a discrete Kleinian boundary.

Before leaving these examples let us consider the cusped manifolds 
\cite{Siegel}\cite{gt3m}\cite{Thur} in some detail. Suppose 
that $X$ is a ($d+1$)-dimensional hyperbolic manifold in either 
of the above examples (thus $d=1,2$), having finite volume $Vol(X)$
and containing $p$ cusps. Mathematically, there is a lower bound 
for the hyperbolic volume:
\begin{equation}
Vol(X)\;\geq\;c_{d}\cdot p
\label{vol-cusps}
\end{equation}
where $c_{d}$ is some constant of order 1, which may depend on what 
the example we are considering, --- in the 2-dimensional example we 
have \cite{Siegel} $c_{1}=2\pi$ and, in the 3-dimensional example 
\cite{Adams}, $c_{2}\approx 1.01494$. According to our geometrical 
construction, the Kleinian boundary $M'$ of $X$ should be 
discretized completely in this finite volume case, consisting of 
the $p$ cusps. The average spacing of points in such a ``lattice 
boundary space'' should behave as 
\begin{equation}
\Delta x_{\parallel}\;\sim\;p^{-1/d}.
\label{spacing}
\end{equation}
Note that (\ref{spacing}) also has an interpretation as the 
uncertainty of localizing a spacetime point in the boundary 
theory. Going to the bulk manifold, the local geometry of $X$ 
reads
\begin{equation}
ds^{2}\;=\;\frac{U^{2}}{R^{2}}\,dx_{\parallel}^{2}\,+\,
\frac{R^{2}}{U^{2}}\,dU^{2},
\label{metric}
\end{equation}
so the volume of $X$ may be estimated roughly
\begin{equation}
Vol(X)\;\sim\;\int^{\Delta U}_{0}dU\,U^{d-1}\;\sim\;(\Delta U)^{d}
\label{vol}
\end{equation}
with $\Delta U$ being an effective infrared cutoff of the bulk
manifold, which can be identified with the length scale of a 
fundamental domain ${\cal F}\subset{\bf H}^{d+1}$ for $\Gamma$. 
Now we combine (\ref{vol-cusps}), (\ref{spacing}) and 
(\ref{vol}) to derive
\begin{equation}
\Delta x_{\parallel}\cdot \Delta U\;\geq\;\sqrt[d]{c_{d}}\;\sim\;1.
\label{UV/IR}
\end{equation}
One recognizes immediately that (\ref{UV/IR}) is the UV/IR 
relation in string units, reformulated as a kind of the 
space-time uncertainty principle \cite{LY}. This gives us further 
evidence that the geometrical definition of $M'$ can serve 
as the space on which the holographic boundary theory lives. The 
discussion also suggests that cusps are capable of storing physical 
information. In appendix A.2, we will give an estimation of 
information density in $d=4$ dimensions.
\section{Conformal Fields on the Boundary}
Having established a mathematical description for the Kleinian 
boundary of $X$, we shall try to explore some unusual behavior of
the boundary conformal fields. In particular we want to show how the 
physical spectrum can depend nontrivially on a generic discrete 
group $\Gamma$. Here, for simplicity, we will only consider 
conformal fields without Lorentz indices; our discussion can be 
easily generalized to the case containing several Lorentz quantum 
numbers. 

We begin with conformal transformations of scalar fields on the 
Kleinian boundar $M'$. For this discussion we need only to consider 
the continuous part $D_{\Gamma}/\Gamma$ of $M'$, ignoring all cusps. 
So given any $f$ in the conformal group (\ref{C-symm}), a scalar field 
${\cal O}(x)$ of conformal dimension $\Delta$ will transform as
\begin{equation}
{\cal O}(x)\;\longrightarrow\;{\cal O}^{f}(x)\;\equiv\;{\cal U}^{-1}_{f}
\,{\cal O}(x)\,{\cal U}_{f}\;=\;|f'(x)|^{\Delta}\,{\cal O}(fx)
\label{tran1}
\end{equation}
where $\cal U$ is some unitary representation of the conformal group and
$|f'(x)|$ is the Euclidean distortion, defined through $|fx-fy|^{2}=
|f'(x)|\cdot |f'(y)|\cdot |x-y|^{2}$ for any $x,y\in\hat{\bf R}^{d}$.

If $\Delta\neq 0$, one cannot really think of ${\cal O}(x)$ as an 
operator-valued distribution since it depends on the choice 
of a Riemannian metric on $D_{\Gamma}/\Gamma$. In particular
under Weyl transformation 
\begin{equation}
h_{ij}\rightarrow \tilde{h}_{ij}\;=\;e^{2w}\cdot h_{ij}
\label{weyl}
\end{equation}
the field ${\cal O}(x)$ will get rescaled, ${\cal O}\rightarrow
\widetilde{\cal O}=e^{-\Delta w}\cdot{\cal O}$. In certain 
applications one may wish to find an intrinsic description for
conformal fields in the sense that it does not depend on the 
choice of Riemannian metrics. For this purpose, we will invoke the
concept of ``conformal densities'' considered in \cite{Sull1}. Recall 
that \cite{Sull1} a conformal density of dimension $\Delta$ is some
positive finite measure $\mu$ assigned to the given metric $h_{ij}$,
and under the Weyl rescaling (\ref{weyl}), it obeys a
transformation law\footnote{Mathematically, this transformation law 
should be understood as a constraint on the two assignments 
$h_{ij}\rightarrow\mu$, $\tilde{h}_{ij}\rightarrow\tilde{\mu}$ 
related by $\tilde{h}_{ij}\;=\;e^{2w}\cdot h_{ij}$, and this 
constraint will force the value of the Radon-Nikodym derivative 
$\frac{d\tilde{\mu}}{d\mu}$ to be $e^{\Delta w}$.}
\begin{equation}
d\mu\rightarrow d\tilde{\mu}\;=\;e^{\Delta w}\cdot d\mu.
\label{density}
\end{equation}
Thus, by taking a measurable subset $A$ of $D_{\Gamma}/\Gamma$, 
one can form an integral invariant under Weyl transformations:
\begin{equation}
{\cal O}(A)\;\equiv\;\int_{A}{\cal O}(x)\,d\mu(x),
\;\;\;\;\;\;A\subset D_{\Gamma}/\Gamma.
\label{intrinsic}
\end{equation}
This clearly gives the desired intrinsic discription for ${\cal O}(x)$. 
Note that the existence of such a description is essential for
formulating the explicit AdS/CFT correspondence \cite{GKP}\cite{Wi1}. 
In that connection, the conformal density $d\mu(x)$ will serve as a 
source coupled to the local field ${\cal O}(x)$ on the boundary. The
coupling should be conformally invariant by physical requirements.

One can lift fields from the quotient manifold $D_{\Gamma}/\Gamma$ to 
its covering space $D_{\Gamma}$ and regard ${\cal O}(x)$ as a conformal 
field $\bar{\cal O}(x)$ on $D_{\Gamma}$, obeying the $\Gamma$-invariant 
condition $U^{-1}_{\gamma}\bar{\cal O}(x)U_{\gamma}=\bar{\cal O}(x)$, 
or, equivalently:
\begin{equation}
\bar{\cal O}(\gamma x)\;=\;|\gamma'(x)|^{-\Delta}\cdot\bar{\cal O}
(x),\;\;\;\;\forall\gamma\in\Gamma.
\label{auto}
\end{equation}
To lift (\ref{intrinsic}), let $\bar{A}$ denote the covering space of 
$A$ and suppose that ${\cal F}\subset D_{\Gamma}$ is a fundamental 
domain for $\Gamma$. Since the images of $\cal F$ under the action of 
$\Gamma$ tesselate $D_{\Gamma}$, we have
\begin{equation}
\begin{array}{lll}
\bar{\cal O}(\bar{A}) & \equiv & {\displaystyle\int_{\bar{A}}
\bar{\cal O}(x)\,d\bar{\mu}(x)\;= \;\sum_{\gamma\in\Gamma}
\int_{\gamma({\cal F}\cap\bar{A})}\bar{\cal O}(x)\,d\bar{\mu}(x)}
\vspace{0.3cm} \\
 & = & {\displaystyle\sum_{\gamma\in\Gamma}\int_{{\cal F}\cap\bar{A}}
 \bar{\cal O}(\gamma x)\,d\bar{\mu}(\gamma x)\;=\;\int_{{\cal F}
 \cap\bar{A}}\bar{\cal O}(x)\,d\nu(x)},
\end{array}
\label{cover}
\end{equation}
where $\bar{\mu}$ is a conformal density on $D_{\Gamma}$, of 
dimension $\Delta$, and
\begin{equation}
d\nu(x)\;=\;\sum_{\gamma\in\Gamma}|\gamma'(x)|^{-\Delta}\cdot 
d\bar{\mu}(\gamma x)
\label{Poincare}
\end{equation}
stands for the $\Gamma$-invariant projection of $d\bar{\mu}(x)$, 
which may be identified with a conformal density on 
$D_{\Gamma}/\Gamma$ having the same dimension as $d\bar{\mu}(x)$. 
Hence the last integral 
in (\ref{cover}) can be regarded as the integral (\ref{intrinsic}) 
under the identifications ${\cal F}\cap\bar{A}\leftrightarrow A$ 
and $d\nu\leftrightarrow d\mu$.

As $D_{\Gamma}$ constitutes a subset of ${\bf S}^{d}$, the 
measure $d\bar{\mu}(x)$ has a natural extension to the whole 
boundary sphere; hence the projection (\ref{Poincare}) defines 
a $\Gamma$-invariant conformal density on ${\bf S}^{d}$. Now 
for generic $\Gamma$, the existence of such an invariant measure 
will impose a constraint on the allowed values of $\Delta$. 
In fact, if $\Gamma$ is a non-elementary discrete group, then 
according to Corollary 4 of \cite{Sull1}, the dimension of any 
conformal density on ${\bf S}^{d}$ invariant under $\Gamma$ has 
a lower bound. This bound can be saturated and agrees with the 
critical exponent $\delta(\Gamma)$ of $\Gamma$, 
which is defined so that the Poincar\'{e} series 
\begin{equation}
g(x,y;s)\;=\;\sum_{\gamma\in\Gamma}\exp(-s\,
\rho(x,\gamma y))\;\;\;\;\;\;
(x,y\in{\bf H}^{d+1},\;\;s\in{\bf R})
\label{poincare}
\end{equation}
converges for $s>\delta(\Gamma)$ and diverges for 
$s<\delta(\Gamma)$, here $\rho(\cdot,\cdot)$ is the 
hyperbolic distance. As a consequence, scalar conformal fields 
on $M'$ are not intrinsically defined in the sense of 
(\ref{intrinsic}) unless their conformal dimensions obey the 
inequality
\begin{equation}
\Delta\;\geq\;\delta(\Gamma).
\label{constraint}
\end{equation}
If there are no stronger constraints presented, $\Delta$ 
can saturate the lower bound in (\ref{constraint}), so in 
this case the conformal dimension of the most relevant operators 
on $M'$ is $\delta(\Gamma)$. 

The above argument rests on the requirement that all conformal 
fields on $D_{\Gamma}/\Gamma$ should have an intrinsic 
description. However, there exists other evidence supporting 
the constraint (\ref{constraint}). Consider, for example, a 
string theory in ${\bf H}^{d+1}\times W$ which corresponds to 
some CFT on the boundary of ${\bf H}^{d+1}$. With a suitable 
choice of $W$ we may assume this string theory contains zero-branes. 
Each zero-brane on the boundary sphere ${\bf S}^{d}$ with mass $m$ 
is associated to a local conformal field ${\cal O}(x)$, whose 
dimension is determined by $\Delta=(d+\sqrt{d^{2}+4m^{2}})/2$. 
The two-point Green's function 
$G(x,y)\equiv\;<{\cal O}(x){\cal O}^{\dagger}(y)>$ 
can be computed within the bulk theory: For $\Delta\sim m\gg 1$, 
we have
\begin{equation}
G(x,y)\;\;\sim\;\;\exp(-m\,\rho(x,y))\;\;\sim\;\;
\exp(-\Delta\,\rho(x,y)).
\label{OO}
\end{equation}
Note that (\ref{OO}) should be regularized by moving $x,y$ 
``slightly'' from the boundary of ${\bf H}^{d+1}$ to its 
interior. With this regularization the Poincar\'{e} series 
$G_{\Gamma}(x,y)\equiv\sum_{\gamma\in\Gamma}G(x,\gamma y)$ 
is biautomorphic, that is, $G_{\Gamma}(\gamma_{1} x,\gamma_{2}y)=
G_{\Gamma}(x,y)$ for any $\gamma_{1},\gamma_{2}\in\Gamma$. 
(This follows from the fact that $\Gamma$ acts isometrically 
on the hyperbolic space, so in particular we have $\rho(\gamma 
x,\gamma y)=\rho(x,y)$, $\forall\gamma\in\Gamma$.) It is therefore 
tampting to interprete $G_{\Gamma}(x,y)$ as a 
regularized two-point Green's function of certain CFT defined 
on the quotient manifold $M'\subset{\bf S}^{d}/\Gamma$. Evidently, 
this interpretation can work only if $G_{\Gamma}(x,y)$ converges. 
When $G_{\Gamma}(x,y)<\infty$, it is possible to construct a local 
conformal field ${\cal O}_{\Gamma}(x)$ on $M'$ such that $<{\cal 
O}_{\Gamma}(x){\cal O}_{\Gamma}^{\dagger}(y)>$ is regularized to 
be $G_{\Gamma}(x,y)$, by a generalized Osterwalder-Schrader 
quantization procedure. The conformal dimension $\Delta$ of such 
an ${\cal O}_{\Gamma}(x)$ should then obey (\ref{constraint}). 
Actually, from (\ref{poincare}) and (\ref{OO}), one easily 
derives 
\begin{equation}
G_{\Gamma}(x,y)\;\;\sim\;\;g(x,y;\,\Delta),
\label{QOO}
\end{equation}
so convergence of $G_{\Gamma}(x,y)$ requires the inequality 
$\Delta\geq\delta(\Gamma)$.

So far we have derived a general constraint (\ref{constraint}) on 
the physical spectrum of conformal fields living in $M'$. While 
this inequality is always presented for conformal fields having 
the intrinsic description (\ref{intrinsic}), it may not really 
provide a useful constraint in the AdS/CFT correspondence. As a 
matter fact, if a scalar conformal field arises from the $AdS_{d
+1}/CFT_{d}$ correspondence, then its conformal dimension should 
obey another inequality \cite{Wi1}
\begin{equation}
\Delta\;\geq\;\frac{d}{2}.
\label{constraint1}
\end{equation}
Thus, if $\delta(\Gamma)\leq\frac{d}{2}$, the constraint 
(\ref{constraint}) is weaker than (\ref{constraint1}) and it 
will be automatically satisfied in the boundary CFT. In this case 
our constraint has no observable effect on the underlying physical 
spectrum. In order for (\ref{constraint}) to have visible effect, 
we have to consider another case, $\delta(\Gamma)>\frac{d}{2}$, 
where taking the quotient by $\Gamma$ will change the boundary theory 
drastically. Note that the case $\delta(\Gamma)>\frac{d}{2}$ is also 
interesting from the mathematical point of view \cite{Sull1}.

The above results can be rederived in the bulk theory via the 
quotient AdS/CFT correspondence. Consider a scalar field 
$\phi(y)$ with mass $m$ in the bulk spacetime $X={\bf H}^{d+1}
/\Gamma$. The classical action is given by 
\begin{equation}
I(\phi)\;=\;\frac{1}{2}\int_{X}d^{d+1}y\,\sqrt{g}
\,(g^{\mu\nu}\,\partial_{\mu}
\phi\,\partial_{\nu}\phi\,+\,m^{2}\phi^{2}). 
\label{action}
\end{equation}
The Klein-Gordon equation for $\phi$ has the form
\begin{equation}
\nabla^{2}\phi\;=\;\lambda\phi
\label{KG}
\end{equation}
with $\nabla^{2}\equiv\frac{1}{\sqrt{g}}\partial_{\mu}
(\sqrt{g}g^{\mu\nu}\partial_{\nu})$ and $\lambda\equiv 
m^{2}$, whose solutions are known as 
$\lambda$-harmonic functions. There is a real number 
\begin{equation}
\lambda_{0}(X)\;=\;-\inf\limits_{\phi}\;
\frac{\int_{X}|\nabla\phi|^{2}}{\int_{X}|\phi|^{2}}\;\leq\;0
\end{equation}
separating the $L^{2}$-spectrum of $\nabla^{2}$ (which is 
contained in the region $-\infty<\lambda\leq\lambda_{0}(X)$) 
from the ``positive'' spectrum (which is contained in the 
region $\lambda_{0}(X)\leq\lambda <\infty$), and for generic 
$\Gamma$ this number takes the value \cite{Sull2}
\begin{equation}
\lambda_{0}(X)\;=\;\left \{
\begin{array}{ll}
{\displaystyle -\frac{1}{4}d^{2},} & {\displaystyle
{\rm if}\;\;\delta(\Gamma)\leq\frac{1}{2}d,}\vspace{.3cm}\\
{\displaystyle\delta(\Gamma)(\delta(\Gamma)-d),} & 
{\displaystyle{\rm if}\;\;\delta(\Gamma)\geq
\frac{1}{2}d.}
\end{array}\right.
\label{EPS}
\end{equation}
Now, given any boundary values $\phi_{0}({\bf x})$ for 
${\bf x}\in M'$ and a positive function $f(y)$ on $\bar{X}
\equiv X\cup M'$ which has a first-order zero on the boundary, 
one wants to extend $\phi_{0}({\bf x})$ uniquely to a 
$\lambda$-harmonic function $\phi$ in the interior of $X$, 
such that $\phi(y)\rightarrow f^{(d-\sqrt{d^{2}+4\lambda})/2}
\cdot\phi_{0}$ as $y\in X$ approaches a boundary point. Stability 
requires that we should only consider positive $\lambda$-harmonic 
extensions, namely solutions of (\ref{KG}) with $\lambda\equiv m^{2}
\geq\lambda_{0}(X)$, and all the $L^{2}$-solutions with $\lambda$ 
strictly less than $\lambda_{0}(X)$ should be discarded (since they 
are precisely the normalizable zero modes that will lead to 
instabilities). According to the AdS/CFT correspondence, the 
conformal operator ${\cal O}({\bf x})$ coupled naturally to the 
source $\phi_{0}({\bf x})$ on the quotient boundary has the scaling 
dimension
\begin{equation}
\Delta\;=\;\frac{d+\sqrt{d^{2}+4m^{2}}}{2}\;\geq\;
\frac{d+\sqrt{d^{2}+4\lambda_{0}(X)}}{2}.
\label{dim}
\end{equation}
Thus, combining this and (\ref{EPS}) we see that $\Delta\geq
\frac{d}{2}$ if $\delta(\Gamma)\leq\frac{d}{2}$ and 
$\Delta\geq\delta(\Gamma)$ if $\delta(\Gamma)\geq\frac{d}{2}$, 
in ageement with the previous argument.

To determine the dependence of $\delta(\Gamma)$ on $\Gamma$, 
let us recall a geometrical characteristic of the critical 
exponent \cite{Sull1}. Suppose first that $\Gamma$ is a convex 
cocompact group, namely, $\Gamma$ has a fundamental domain in 
${\bf H}^{d+1}$ which has finitely many sides and does 
not touch the limit set $\Lambda_{\Gamma}$. Sometimes such 
groups are also called ``geometrically finite without cusps'' 
and they form quite a rich class in hyperbolic geometry. If 
$\Gamma$ is convex cocompact, then it can be shown (\cite{Sull1}, 
Section 3) that the critical exponent of $\Gamma$ agrees 
exactly with the Hausdorff dimension of the limit set:
\begin{equation}
\delta(\Gamma)\;=\;d_{H}(\Lambda_{\Gamma}).
\label{theorem7}
\end{equation}

Next, for a more general non-elementary discrete group $\Gamma$, 
its critical exponent turns out to be greater than or equal to 
the Hausdorff dimension of the radial limit set 
$\Lambda_{\Gamma}^{rad}$. Recall that \cite{Sull1} a 
point $x$ of ${\bf S}^{d}$ belongs to $\Lambda_{\Gamma}^{rad}$ if and 
only if there are infinitely many $y_{1},y_{2},\dots\in{\bf H}^{d+1}$ 
in some orbit $\Gamma\cdot y$, such that the distance between $y_{n}$ 
and a geodesic ray ending at $x$ is bounded for all $n$. Using this 
definition one easily checks that $\Lambda_{\Gamma}^{rad}\subset
\Lambda_{\Gamma}$. It is conjectured that not only the inequality 
$\delta(\Gamma)\geq d_{H}(\Lambda_{\Gamma}^{rad})$ holds, but actually 
one has the equality
\begin{equation}
\delta(\Gamma)\;=\;d_{H}(\Lambda_{\Gamma}^{rad}),
\label{theorem25}
\end{equation}
and this conjecture was verified when $\Gamma$ has a $\delta(
\Gamma)$-finite volume (\cite{Sull1}, Section 6), including all 
(finitely generated) Fuchian groups. Note that (\ref{theorem25}) 
implies (\ref{theorem7}) in the convex cocompact case since in 
that case we have $\Lambda_{\Gamma}^{rad}=\Lambda_{\Gamma}$.

From the foregoing discussion we expect that $\delta(\Gamma)$ 
may increase as $\Gamma$ becomes larger and larger. In particular 
when $\Lambda_{\Gamma}^{rad}=\Lambda_{\Gamma}$ fills the whole 
boundary sphere ${\bf S}^{d}$, or the Hausdorff dimension 
$d_{H}(\Lambda_{\Gamma}^{rad})$ reaches its maximal value 
$d$, then according to the constraint (\ref{constraint}), there 
will be no relevant operators in the boundary theory. As we have 
seen in the previous section, this can occur if $\Gamma$ becomes 
so large that the volume of $X$ is finite; in such an extreme case 
we have $m^{2}\geq\lambda_{0}(X)=0$ for all scalar fields in the 
bulk and thus no tachyon modes can exist. It would be very 
interesting to understand the underlying bulk/boundary correspondence
in terms of brane dynamics (cf. \cite{probe}).
\section{Computing the Scalar 2-point Function}
In this section we shall compute the scalar two-point function 
explicitly using the quotient AdS/CFT correspondence. The 
computation is a $\Gamma$-invariant version of that presented 
in \cite{Wi1}. For simplicity, we shall assume that $\Gamma$ has 
no cusps, or at least the cusps do not meet the limit set 
$\Lambda_{\Gamma}$, so that the Kleinian boundary of 
$X={\bf H}^{d+1}/\Gamma$ is simply given by $M'=D_{\Gamma}/\Gamma$.

Following \cite{Wi1}, let us solve the Klein-Gordon equation 
(\ref{KG}) defined in $X$ with the given boundary values 
$\phi_{0}({\bf x})$, ${\bf x}\in M'$. For this purpose we need 
to construct a (generalized) Poisson kernel $K(y,{\bf x})$ on 
$X\times M'$ such that $(\nabla^{2}_{y}-m^{2})K(y,{\bf x})=0$ 
for all $y$ in $X$ and 
\begin{equation}
\phi(y)\;=\;\int_{M'}d^{d}{\bf x}\,K(y,{\bf x})
\cdot\phi_{0}({\bf x}) 
\label{phi}
\end{equation}
has the correct boundary behavior. On the covering space 
${\bf H}^{d+1}\times D_{\Gamma}$ of $X\times M'$, the Poisson 
kernel $k(y,{\bf x})$ is known to be
\cite{Wi1}
\begin{equation}
k(y,{\bf x})\;=\;const.\times 
\frac{y_{0}^{\Delta}}{(y_{0}^{2}+|{\bf y}
-{\bf x}|^{2})^{\Delta}}
\label{ckernel}
\end{equation}
where $\Delta$ is defined in (\ref{dim}). However, this 
expression does not meet our requirement of $\Gamma$-invariance. 
In fact, given any $\gamma\in\Gamma$, we have 
$\rho(\gamma y,\gamma x)=\rho(y,x)$ and $\cosh\rho(y,x)
=1+\frac{|y-x|^{2}}{2y_{0}x_{0}}$, so that
\begin{equation}
k(\gamma y,\gamma {\bf x})\;=\;
(\lim_{x_{0}\rightarrow 0}\frac{x_{0}}{
\gamma(x)_{0}})^{\Delta}\cdot k(y,{\bf x})
\;=\;|\gamma'({\bf x})|^{-\Delta}
\cdot k(y,{\bf x}).
\label{transform}
\end{equation}
Thus in general we do not have 
$k(\gamma y,{\bf x})=k(y,{\bf x})$. To 
find the $\Gamma$-invariant Poisson kernel, 
we will invoke the familiar Poincar\'{e} series
\begin{equation}
K(y,{\bf x})\;\equiv\;\sum_{\gamma\in\Gamma}
\,|\gamma'({\bf x})|^{\Delta}\cdot 
k(y,\gamma{\bf x}),\;\;\;\;\;\;y\in{\bf H}^{d+1},
\;\;{\bf x}\in D_{\Gamma}.
\label{kernel}
\end{equation}
It is easy to check that (i) $K(y,{\bf x})$ has the desired 
$\Gamma$-invariant propertiy 
$K(\gamma y,{\bf x})=K(y,{\bf x})$ for any
$\gamma\in\Gamma$, (ii) $K(y,{\bf x})$ transforms 
covariantly under ${\bf x}\rightarrow \gamma{\bf x}$, 
namely $K(y,\gamma{\bf x})=|\gamma'({\bf x})|^{-\Delta}
\cdot K(y,{\bf x})$, and (iii) $K(y,{\bf x})$ solves 
the Klein-Gordon equation in ${\bf H}^{d+1}$. Thus, 
combining (i) and (iii) we see that (\ref{kernel}) is in fact 
a solution of the Klein-Gordon equation in $X$. Before we can 
identify such $K(y,{\bf x})$ with the Poisson kernel defined on 
$X\times M'$, we need to check further that substituting 
(\ref{kernel}) into (\ref{phi}) will lead to the correct 
boundary behavior of $\phi(y)$.

To this end we take a $\Gamma$-invariant conformal density 
(in the sense of \cite{Sull1}) of dimension $\Delta$, defined 
on the covering space $D_{\Gamma}$ of the quotient boundary:
\begin{equation}
d\bar{\mu}({\bf x})\;\equiv\;
\bar{\phi}_{0}({\bf x})\,d^{d}{\bf x},\;\;\;\;\;\;
d\bar{\mu}(\gamma{\bf x})\;=\;|\gamma'({\bf x})|^{\Delta}\cdot
d\bar{\mu}({\bf x}).
\label{density1}
\end{equation}
One requires that the function $\bar{\phi}_{0}({\bf x})$ 
in (\ref{density1}) should transform as 
$\bar{\phi}_{0}(\gamma {\bf x})\;=\;
|\gamma'({\bf x})|^{\Delta -d}\cdot\bar{\phi}_{0}({\bf x})$ 
under $\gamma\in\Gamma: D_{\Gamma}\rightarrow D_{\Gamma}$, 
namely, it should look like a field of conformal 
dimension $d-\Delta$. In particular we can take 
$\bar{\phi}_{0}({\bf x})$ to be the lifting of the boundary 
values ${\phi}_{0}({\bf x})$ from $M'$ to its covering space 
$D_{\Gamma}$, which has the correct conformal dimension. 
With this choice we pick an arbitrary fundamental domain 
${\cal F}\subset D_{\Gamma}$ of $\Gamma$ and then consider
\begin{equation}
\begin{array}{lll}
{\displaystyle\int_{D_{\Gamma}}d^{d}{\bf x}
\,k(y,{\bf x})\,\bar{\phi}_{0}({\bf x})}
& = & {\displaystyle\sum_{\gamma\in\Gamma}
\int_{\gamma({\cal F})}
k(y,{\bf x})\,d\bar{\mu}({\bf x})}
\;=\;{\displaystyle\sum_{\gamma\in\Gamma}
\int_{\cal F} k(y,\gamma{\bf x})
\,|\gamma'({\bf x})|^{\Delta}\, d\bar{\mu}({\bf x})}
\vspace{0.3cm} \\
 & = & {\displaystyle\int_{{\cal F}}d^{d}
 {\bf x}\,K(y,{\bf x})\,\bar{\phi}_{
 0}({\bf x})\;=\;\int_{M'}d^{d}{\bf x}\,K(y,{\bf x})
 \,\phi_{0}({\bf x})}.
\end{array}
\label{integral}
\end{equation}
Notice that the first integral in (\ref{integral}) 
has the correct boundary behavior, since $k(y,{\bf x})$ 
is the Poisson kernel on the covering space. The last integral 
thus gives the desired solution of the Klein-Gordon equation 
in $X$ with the boundary values $\phi_{0}({\bf x})$.

The above discussion also explains why the boundary theory dual to
supergravity on $X$ should live exactly on $M'$. If it is
defined on a subspace $N$ of $M'$ with $M'-N\neq\emptyset$, then
we can cover $N$ by a subspace $E$ of $D_{\Gamma}$ such that
$D_{\Gamma}-E$ is non-empty and invariant under $\Gamma$. We may
thus replace $\cal F$ in (\ref{integral}) by ${\cal F}\cap E$ to
derive
\begin{equation}
\int_{E}d^{d}{\bf x}\,k(y,{\bf x})\,\bar{\phi}_{0}({\bf x})
\;=\;\int_{N}d^{d}{\bf x}\,K(y,{\bf x})\,\phi_{0}({\bf x})\;\equiv
\;\phi(y),
\label{integral1}
\end{equation}
where $\phi(y)$ solves the equations of motion in the bulk. However,
this solution has a rather special asymptotic behavior as $y$ in the
l.h.s of (\ref{integral1}) approaches $D_{\Gamma}-E$, and
the ``boundary value'' $\bar\phi_{0}(y)\equiv y_{0}^{\Delta-d}\phi(y)$
vanishes at any $y={\bf y}\in D_{\Gamma}-E$ due to the known
behavior of the Poinsson kernel on the covering space \cite{Wi1},
$k({\bf y},{\bf x})\propto y_{0}^{d-\Delta}\delta^{d}({\bf x}-{\bf y})$.
Thus (\ref{integral1}) cannot describe more general solutions
with the asymptotic behavior $\lim\limits_{y\rightarrow{\bf y}}
\phi_{0}(y)\neq 0$ for ${\bf y}\in (D_{\Gamma}-E)/\Gamma$, which
is of course allowed in the quotient bulk theory. It follows that
bulk supergravity on $X$ cannot be faithfully represented by a
theory on $N\subset M'$, unless $N=M'$. On the other hand, if the
boundary theory is defined on a space $N$ containing $M'$ with
$N-M'\neq\emptyset$, then the covering space $E$ of $N$ must consist
of $D_{\Gamma}$ together with some points (not cusps) in the limit set
$\Lambda_{\Gamma}$. When this happens, Eq.(\ref{integral1}) will no
longer hold, since $\Gamma$ now does not act properly discontinuously
on $E$ and accordingly we cannot decompose $E$ canonically into
fundamental domains ${\cal F}\cap E$. To construct a solution of the
Klein-Gordon equation in $X$ using boundary data on $N=E/\Gamma$,
the best we can do is to formulate the integral (\ref{integral}) over
$D_{\Gamma}$. One may then ask whether the $\Gamma$-invariant
boundary data $\bar{\phi}_{0}({\bf x})$ for $\bf x$ outside $D_{\Gamma}$
is redundant. The answer is affirmative because each ${\bf x}\in
E-D_{\Gamma}$ is determined by ${\bf x}=\lim\limits_{n\rightarrow\infty}
\gamma_{n}({\bf x}')$ with ${\bf x}'\in D_{\Gamma}$ and $\gamma_{n}\in
\Gamma$, so as long as the boundary values of $\phi$ on $D_{\Gamma}/
\Gamma$ are given, we need not to specify the values of $\phi$
separately on $(E-D_{\Gamma})/\Gamma$. In other words, the space
$M'=D_{\Gamma}/\Gamma$ is sufficient to define the boundary conditions
for solving the Klein-Gordon equation in $X$.

We are now ready to evaluate the action (\ref{action}). 
Let us choose a fundamental domain ${\cal G}$ of $\Gamma$ 
in ${\bf H}^{d+1}$ such that $\bar{\cal G}\cap{\bf S}^{d}
=\bar{\cal G}\cap D_{\Gamma}={\cal F}$. This is 
possible since we have assumed that all possible cusps 
of $\Gamma$ do not live in the limit set $\Lambda_{\Gamma}$. 
We can represent the integral in (\ref{action}) over $X$ by a 
corresponding integral over $\cal G$ and, by integrating by parts, 
one finds that the action $I(\phi)$ consists of two terms, one of 
which vanishes due to the equations of motion and the other of 
which can be reduced to a surface integral \cite{Wi1}
\begin{equation}
I(\phi)\;=\;\frac{1}{2}\int_{\cal G}d^{d+1}y\,
\partial_{\mu}\,(\sqrt{g}\,g^{\mu\nu}\,\phi
\,\partial_{\nu}\phi)\;=\;\lim_{\epsilon\rightarrow 0}
\;\frac{1}{2}\int_{\bar{\cal G}\cap T_{\epsilon}} 
d^{d}{\bf y}\,\sqrt{h}\,
\phi\,{\bf n}\cdot\nabla\,\phi
\label{action1}
\end{equation}
where $T_{\epsilon}$ is the horosphere $y_{0}=\epsilon$ 
based at $\infty$, $h_{ij}=\delta_{ij}/\epsilon^{2}$ is 
the metric induced from $g_{\mu\nu}$, and ${\bf n}\cdot\nabla
=y_{0}\frac{\partial}{\partial y_{0}}|_{y_{0}=\epsilon}$
is the derivative along the direction normal to the suface 
$\bar{\cal G}\cap T_{\epsilon}$ in $\cal G$. Now using 
(\ref{phi}), (\ref{ckernel}), (\ref{kernel}) and the fact 
$\lim\limits_{\epsilon\rightarrow 0}\bar{\cal G}
\cap T_{\epsilon}={\cal F}$, one gets
\begin{equation}
I(\phi)\;=\;const.\times\int_{{\cal F}\times{\cal F}}
d^{d}{\bf x}d^{d}{\bf y}\,\bar{\phi}_{0}({\bf x})
\left (\sum_{\gamma\in\Gamma}\frac{|\gamma'({\bf y})|^{
\Delta}}{|{\bf x}-\gamma{\bf y}|^{2\Delta}}
\right )\bar{\phi}_{0}({\bf y})
\label{action2}
\end{equation}
with $\bar{\phi}_{0}({\bf x})$ being the lifting of 
$\phi_{0}({\bf x})$.

Since our computation manifests $\Gamma$-invariance, the 
above expression is actually an integral over $M'\times M'$. 
This may be seen directly from (\ref{action2}). For this 
purpose, we take two M\"{o}bius transformations 
${\bf x}\rightarrow\gamma_{1}{\bf x}$, ${\bf y}
\rightarrow\gamma_{2}{\bf y}$. 
Using (\ref{density1}) as well as the distortion formula
\begin{equation}
\frac{|\gamma'(\gamma_{2}{\bf y})|^{\Delta}}
{|\gamma_{1}{\bf x}-\gamma\gamma_{2}{\bf y}|^{2\Delta}}
\;=\;\frac{|\gamma'_{1}({\bf x})|^{-\Delta}\cdot|(\gamma_{1}^{-1}
\gamma\gamma_{2})'({\bf y})|^{\Delta}
\cdot|\gamma'_{2}({\bf y})|^{-\Delta}}
{|{\bf x}-\gamma^{-1}_{1}\gamma\gamma_{2}{\bf y}|^{2\Delta}}
\label{distortion}
\end{equation}
(which may be verified directly from the definition), 
we find that the integral (\ref{action2}) is invariant 
if we replace its integration domain ${\cal F}\times{\cal F}$ 
by $\gamma_{1}({\cal F})\times\gamma_{2}({\cal F})$ 
and, at the same time, replace $\Gamma$ by 
$\tilde{\Gamma}\equiv\gamma_{1}\Gamma\gamma_{2}^{-1}$. 
Thus, if $\gamma_{1},\gamma_{2}\in\Gamma$, then 
$\tilde{\Gamma}$ coincides exactly with $\Gamma$ and 
$\gamma_{1,2}({\cal F})$ constitute fundamental domains. 
Consequently, (\ref{action2}) is independent of the choice 
of fundamental domains in both factors of 
${\cal F}\times{\cal F}$, hence giving rise to an 
integral of the form 
\begin{equation}
I(\phi)\;=\;\int_{M'\times M'}d^{d}{\bf x}d^{d}{\bf y}\,
\phi_{0}({\bf x})\,G({\bf x},{\bf y})\,\phi_{0}({\bf y}),\;\;\;\;
G({\bf x},{\bf y})\;\propto\;\sum_{\gamma\in\Gamma}\frac{|\gamma'
({\bf y})|^{\Delta}}{|{\bf x}-\gamma{\bf y}|^{2\Delta}}.
\label{action3}
\end{equation}

According to the quotient version of the AdS/CFT 
correspondence, the function $G({\bf x},{\bf y})$ 
in the bulk action (\ref{action3}) is to be identified 
with the two-point correlator $<{\cal O}({\bf x}){\cal O}
({\bf y})>$ of the boundary theory, where ${\cal O}({\bf x})$ 
is a scalar field on $M'$ of conformal dimension $\Delta$. 
We thus expect that $G({\bf x},{\bf y})$ should be consistent 
with conformal invaiance of the boundary theory; in 
particular, it should transform correctly under the 
unbroken conformal group (\ref{C-symm}) on $M'$, namely 
for any $f\in SO(d+1,1)$ such that $f^{-1}\Gamma 
f=\Gamma$, we should have $G(f({\bf x}),f({\bf y}))
=|f'({\bf x})|^{-\Delta}|f'({\bf y})|^{-\Delta}G({\bf x},
{\bf y})$. As a quick check, taking $\gamma_{1}
=\gamma_{2}=f$ in (\ref{distortion}) to derive
\begin{equation}
\sum_{\gamma\in\Gamma}\frac{|\gamma'(f({\bf y}))|^{\Delta}}{|f({\bf x})
-\gamma f({\bf y})|^{2\Delta}}\;=\;|f'({\bf x})
|^{-\Delta}|f'({\bf y})|^{-\Delta}\sum_{\gamma\in f^{-1}\Gamma f}
\frac{|\gamma'({\bf y})
|^{\Delta}}{|{\bf x}-\gamma {\bf y}|^{2\Delta}}.
\label{green}
\end{equation}
Hence $G({\bf x},{\bf y})$ indeed has the desired 
transformation properties under (\ref{C-symm}).

The above computation has several straightforward extensions.
First we can consider ``twisted sectors'' by incorporating a 
factor $\chi(\gamma)$ in each term of the Poincar\'{e} series 
(\ref{kernel}), where $\chi$ is some one-dimensional unitary 
representation of the discrete group $\Gamma$. Secondly, the
computation can be extended directly to correlation functions 
involving other local fields in the boundary theory, such as 
fields with Lorentz and internal indices. The general form of 
two-point functions should look like a Poincar\'{e} series which 
arises from taking the $\Gamma$-invariant projection. It is also 
possible to compute higher-point functions by investigating 
nontrivial interactions in the supergravity action.
\section{Conclusions}
In this paper we have made several observations supporting a 
quotient version of the AdS/CFT correspondence. One observation 
is that given bulk spacetime of the form $X={\bf H}^{d+1}/\Gamma$, 
the quantized conformal field theory associated to tree-level 
supergravity in $X$ should live on the Kleinian boundary 
$M'=(D_{\Gamma}/\Gamma)\cup\{cusps\;not\;in\;D_{\Gamma}\}$ of
$X$, rather than on the naive quotient space 
$M={\bf S}^{d}/\Gamma$. This is consistent both with the 
holographic principle and with a geometrical consideration. 
In particular, we explained how this Kleinian boundary manifold 
can be completely discretized when the domain of discontinuity 
$D_{\Gamma}$ collapses or, equivalently, the limit set 
$\Lambda_{\Gamma}$ fills the whole boundary sphere 
${\bf S}^{d}$ of the hyperbolic space, with its Hausdorff 
dimension $d_{H}(\Lambda_{\Gamma})$ reaching the maximal 
value $d$. In this extreme case we found an interesting
physical interpretation of a mathematical relation between 
the volume of cusped hyperbolic manifolds and the number of 
cusps on them, in terms of the space-time uncertainty principle.

In order for conformal fields on $M'$ to have an intrinsic 
description (required for formulating the AdS/CFT 
correspondence), we argued that the spectrum of conformal 
dimensions should be bounded from below by a critical 
exponent $\delta(\Gamma)$, which is roughly the same as the 
Hausdorff dimension $d_{H}(\Lambda_{\Gamma})$ of the limit 
set. We have given two arguments, one which was based on the 
boundary theory and the other rested on bulk supergravity, both 
led to the same bound. This bound will have visible effect 
provided it exceeds another bound, $d/2$, established in spaces 
without taking the quotient. The critical exponent 
$\delta(\Gamma)$ gives a natural measure of the ``size'' of 
$\Gamma$, so when $\Gamma$ becomes sufficiently large the 
scaling dimension of the most relevant operators will
increase. In the extreme case $\delta(\Gamma)=d$ we could 
have a boundary theory without relevant operators.

We also computed the scalar two-point function by reducing 
the action of massive scalars in $X$ to a bilinear term on 
the quotient boundary $M'$. The result turned out to be 
consistent with conformal invariance of the boundary theory. 
In this computation, we have restricted ourselves to the simplest 
case $M'=D_{\Gamma}/\Gamma$ and ignored the possibility that 
cusps of $\Gamma$ can live outside the domain of discontinuity. 
If there are cusps touching the limit set, then our computation 
would be complicated by possible contributions from the
``cusp forms''. In general, these contributions are expected to 
be not negligible since the hand-wave argument presented at the 
end of section 3 suggests that a neighbourhood of a cusp, even it 
is very small, could contain nontrivial physical information. In a 
sense, cusps look like black holes, both of which render 
the manifold incomplete. It would be interesting to know more 
about the role of cusps in the AdS/CFT correspondence.
\section *{Acknowledgments}
I would like to thank C. Xiong, M. Yu, C. Zhu, and W. Zhang
for discussions.
\section *{Note Added}
After this paper was submitted as e-print, I became aware of an existing
work \cite{Kh} where a similar connection between the AdS/CFT 
correspondence and geometry and topology of hyperbolic manifolds was
observed. Closely related results were also reported in the earlier
work \cite{HM}, in which the quotient generalization of the AdS/CFT 
duality was first proposed. I would like to thank G. Horowitz for 
pointing out the reference \cite{HM} to me.
\appendix
\section{Appendix}
In this appendix, we will give further clarification of some issues
addressed in the main body of this paper.
\subsection{$\Gamma$-invariant Truncating}
At the beginning of section three, we mentioned that in the quotient 
generalization of the AdS/CFT correspondence, the boundary theory 
associated to supergravity on $X$ should be identified with a 
$\Gamma$-invariant truncation (or projection) of the usual boundray 
CFT, dual to the original AdS supergravity without taking the 
quotient. We now want to show that this is indeed a reasonable
speculation. We will first consider $\Gamma$-invariant projection of 
the usual boundary CFT in some detail, and then compare it to the 
boundary theory explored in the text.

For the usual boundary CFT defined on ${\bf S}^{d}$, a scalar operator 
$\widehat{\cal O}({\bf x})$ of conformal dimension $\Delta$ transforms 
under $f\in SO(d+1,1)$ as 
\begin{equation}
\widehat{\cal O}({\bf x})\;\longrightarrow\;\widehat{\cal O}^{f}(
{\bf x})\;\equiv\;{\cal U}^{-1}_{f}\,\widehat{\cal O}({\bf x})\,
{\cal U}_{f}\;=\;|f'({\bf x})|^{\Delta}\,\widehat{\cal O}(f{\bf x}),
\label{tran2}
\end{equation}
which has the familiar two-point correlation function
\begin{equation} 
<\widehat{\cal O}({\bf x})\widehat{\cal O}({\bf y})>\;=\;
\frac{const.}{|{\bf x}-{\bf y}|^{2\Delta}}.
\label{2-point}
\end{equation}
Consider now the $\Gamma$-invariant projection of $\widehat{\cal O}
({\bf x})$ defined by the Poincar\'{e} series:
\begin{equation}
{\cal O}({\bf x})\;=\;\sum_{\gamma\in\Gamma}\,
|\gamma'({\bf x})|^{\Delta}\,\widehat{\cal O}(\gamma{\bf x}).
\label{proj}
\end{equation}
Under the action of a conformal transformation $f\in SO(d+1,1)$, the 
change in the projected field (\ref{proj}) will be computed according
to (\ref{tran2}):
\begin{equation}
\begin{array}{lll}
{\cal O}({\bf x})\;\rightarrow\;{\cal O}^{f}({\bf x}) & \equiv &
{\displaystyle{\cal U}^{-1}_{f}\,{\cal O}({\bf x})\,{\cal U}_{f}
\;=\;\sum_{\gamma\in\Gamma}\,|\gamma'({\bf x})|^{
\Delta}\,{\cal U}^{-1}_{f}\,\widehat{\cal O}
(\gamma{\bf x})\,{\cal U}_{f}} 
\vspace{0.3cm} \\
 & = & {\displaystyle\sum_{\gamma\in\Gamma}\,|f'(\gamma{\bf x})|^{
\Delta}\cdot |\gamma'({\bf x})|^{\Delta}\,\widehat{\cal O}(f\gamma
{\bf x})}
\vspace{0.3cm} \\
 & = & {\displaystyle\sum_{\gamma\in\Gamma}\,|(f\gamma)'(
{\bf x})|^{\Delta}\,\widehat{\cal O}(f\gamma{\bf x})}
\vspace{0.3cm} \\
 & = & {\displaystyle |f'({\bf x})|^{\Delta}
 \sum_{\gamma\in\Gamma}\,|(f\gamma f^{-1})'(f{\bf x})|^{\Delta}\,
 \widehat{\cal O}(f\gamma f^{-1}\cdot f{\bf x})}
\vspace{0.3cm} \\
 & = & {\displaystyle |f'({\bf x})|^{\Delta}\sum_{\gamma\in 
f\Gamma f^{-1}}\,|\gamma'(f{\bf x})|^{\Delta}\,\widehat{\cal O}
(\gamma f{\bf x})}.
\end{array}
\label{tran3}
\end{equation}

The above computation tells us two things: First, if $f\in\Gamma$, 
then $f\gamma\in\Gamma$ and we can rearrange the sum in the third line 
of (\ref{tran3}) to derive ${\cal O}^{f}({\bf x})={\cal O}({\bf x})$, 
which clearly indicates that the projection (\ref{proj}) indeed 
defines a $\Gamma$-invariant field. Second, given $f\in SO(d+1,1)$, 
the last identity in (\ref{tran3}) shows that the projected operator 
${\cal O}({\bf x})$ will transform as a conformal field of dimension 
$\Delta$
\begin{equation}
{\cal O}({\bf x})\;\rightarrow\;{\cal O}^{f}({\bf x})\;=\;
|f'({\bf x})|^{\Delta}{\cal O}(f{\bf x})\;=\;|f'({\bf x})
|^{\Delta}\sum_{\gamma\in\Gamma}\,|\gamma'(f{\bf x})|^{\Delta}\,
\widehat{\cal O}(\gamma f{\bf x})
\label{tran4}
\end{equation}
if and only if $f\Gamma f^{-1}=\Gamma$. This implies that performing
the projection (\ref{proj}) will break the full conformal group 
$SO(d+1,1)$ to a subgroup defined by (\ref{C-symm}). Thus, on symmetry 
grounds, the $\Gamma$-invariant truncation of the usual boundary CFT
should agree with the boundary theory dual to supergravity on $X$, as
both theories have the same conformal group. 

In addition to this, the scalar two-point function of the truncated 
theory can be evaluated by substituting (\ref{proj}) into 
(\ref{2-point}):
\begin{equation}
<{\cal O}({\bf x}){\cal O}({\bf y})>\;\propto\;\sum_{\gamma_{1},
\gamma_{2}\in\Gamma}\,\frac{|\gamma'_{1}({\bf x})|^{\Delta}|\gamma'_{2}
({\bf y})|^{\Delta}}{|\gamma_{1}{\bf x}-\gamma_{2}{\bf y}|^{2\Delta}}
\;=\;\sum_{\gamma_{1},\gamma_{2}\in\Gamma}\,\frac{|(\gamma_{1}^{-1}
\gamma_{2})'({\bf y})|^{\Delta}}{|{\bf x}-\gamma_{1}^{-1}\gamma_{2}
{\bf y}|^{2\Delta}}
\label{proj2}
\end{equation}
Thus $<{\cal O}({\bf x}){\cal O}({\bf y})>$ coincides with the two-point 
function $G({\bf x},{\bf y})$ derived in (\ref{action3}), up to an 
overall infinite factor $|\Gamma|$ that can be absorbed by normalization 
of the vacuum state in the truncated theory. One expects that such a 
coincidence also holds for higher point functions.

The foregoing discussion suggests that the boundary theory studied in the
text is not very different from the boundary CFT dual to the usual 
AdS supergravity; it is likely that the former can be constructed 
by $\Gamma$-invariant truncating of the latter. The only subtleties
in this construction may arise as whether the projection (\ref{proj}) 
or its ``topological dual'' is a mathematically well-defined object. 
Since ${\cal U}_{\gamma}{\cal O}({\bf x}){\cal U}^{-1}_{\gamma}=
{\cal O}({\bf x})$ for any $\gamma\in\Gamma$, naively the operator 
(\ref{proj}) would be defined on the orbit space $M={\bf S}^{d}/\Gamma$; 
but as we saw in the main body of this paper, $M$ in general is not a 
Hausdorff space and in order to formulate a reasonable boundary CFT, 
we have to restrict ourselves to the subspace $M'\subset M$. Moreover, 
in the usual AdS/CFT correspondence the scalar operator $\widehat{\cal O}
({\bf x})$ on ${\bf S}^{d}$ is associated to some conformal density 
$d\widehat{\mu}$ of dimension $\Delta$ through the conformally invariant 
coupling $\int \widehat{\cal O}({\bf x}) d\widehat{\mu}({\bf x})$. When 
one extends this coupling to the truncated operator (\ref{proj}), one has 
to consider the integral
\begin{equation}
\int {\cal O}({\bf x}) d\widehat{\mu}({\bf x})\;=\;\int
\sum_{\gamma\in\Gamma}\,|\gamma'({\bf x})|^{\Delta}
\widehat{\cal O}(\gamma{\bf x}) d\widehat{\mu}({\bf x})\;=\;
\int\widehat{\cal O}({\bf y})\sum_{\gamma\in\Gamma}
|(\gamma^{-1})'({\bf y})|^{-\Delta} d\widehat{\mu}(\gamma^{-1}{\bf y}),
\label{proj3}
\end{equation}
where we have used the identity $|\gamma'(\gamma^{-1}{\bf y})|\cdot
|(\gamma^{-1})'({\bf y})|=1$. This forces us to consider the 
$\Gamma$-invariant projection $d\mu({\bf x})=\sum_{\gamma\in\Gamma}
|\gamma'({\bf x})|^{-\Delta} d\widehat{\mu}(\gamma{\bf x})$ of the 
conformal density, whose existence requires the condition $\Delta\geq
\delta(\Gamma)$ as described in section 4. Thus, modulo these subtleties, 
the $\Gamma$-invariant truncation of the usual CFT dual to supergravity on
${\bf H}^{d+1}$ should be identified with the boundary theory associated
to the bulk theory on $X$.
\subsection{Information Density in the Boundary}
We turn now to an estimation of information density in the boundary CFT
dual to supergravity on the quotient bulk $X$. For simplicity, let us 
restrict ourselves to $d=4$ dimensions; we have just seen that the 
underlying boundary theory on $M'$ can be regarded as the $\Gamma$-invariant 
truncation of the usual super Yang-Mills theory defined on ${\bf S}^{4}$. 
This observation allows us to estimate the number of degrees of freedom 
in a way similar to that presented in \cite{SW}. As in section 5, we 
shall focus mainly on the case where cusps of $\Gamma$ do not touch the 
limit set, so that $M'=D_{\Gamma}/\Gamma$. Contributions from cusps to the 
amount of information may be investigated along the line of establishing 
the space-time uncertainty relation (\ref{UV/IR}); see also the ending
paragraph of this appendix for a related discussion. The results derived
below might have a generalization to other dimensions as well, by 
incorporating the analysis given in \cite{PP}.

Locally, the hyperbolic metric on $X={\bf H}^{5}/\Gamma$ can be written 
in the form
\begin{equation}
ds^{2}\;=\;\frac{R^{2}}{x_{0}^{2}}\,(dx_{0}^{2}+|d{\bf x}|^{2}),\;\;\;\;
R\equiv l_{s}(2\pi g_{s}N)^{1/4},\;\;\;\;{\bf x}\equiv(T,X_{1},X_{2},X_{3})
\label{metric1}
\end{equation}
where the dimensionless coordinates $(x_{0},{\bf x})$ describe points 
of a certain fundamental domain ${\cal G}\subset{\bf H}^{5}$ for 
$\Gamma$. Let $\bar{\cal G}$ denote the closure of $\cal G$. The set 
$\bar{\cal G}\cap D_{\Gamma}$ then gives a fundamental domain for 
$\Gamma$ living in $D_{\Gamma}$, which can be identified with the 
quotient boundary $M'$. Since we are considering the nontrivial case 
where the boundary set has a non-zero measure, the hyperbolic volume 
of $X$ diverges and thus the bulk theory needs to be regulated. One 
simple way to do so is to take the horosphere $T_{\epsilon}$ with 
$\epsilon \ll 1$ as in (\ref{action1}), replacing the boundary set 
$\bar{\cal G}\cap D_{\Gamma}$ by $\bar{\cal G}\cap T_{\epsilon}$. The
resulting space has the induced metric
\begin{equation}
ds^{2}\;=\;\frac{R^{2}}{\epsilon^{2}}\,(dT^{2}+dX_{1}^{2}+dX_{2}^{2}
+dX_{3}^{2}).
\label{metric2}
\end{equation}
Consequently, the spacial area on the boundary set is regulated by
\begin{equation}
A\;=\;\frac{R^{3}}{\epsilon^{3}}\,\int dX_{1}\,dX_{2}\,dX_{3}\;\approx\;
\frac{R^{3}}{\epsilon^{3}}\,(\Delta U)^{3}
\label{area}
\end{equation}
where $\Delta U$ denotes the (dimensionless) Euclidean length scale of 
the regulated space $\bar{\cal G}\cap T_{\epsilon}$. The quantity 
$1/\epsilon$ may be thought of as an IR cutoff of the bulk since the 
hyperbolic volume of $X_{\epsilon}\equiv\{x\in X|x_{0}>\epsilon\}$ 
becomes finite 
\begin{equation}
Vol(X_{\epsilon})\;=\;R^{5}\int_{\epsilon}^{\infty}\frac{dx_{0}}{x_{0}^{5}}
\,\int_{\bar{\cal G}\cap T_{\epsilon}} dT\,dX_{1}\,dX_{2}\,dX_{3}\;\approx\;
R^{5}\,\left(\frac{\Delta U}{\epsilon}\right)^{4},
\label{volume}
\end{equation}
and the
boundary of $X_{\epsilon}$ can be identified with $\bar{\cal G}\cap 
T_{\epsilon}$. 

We are now ready to estimate the information density. To test holography in
the quotient generalization of the AdS/CFT correspondence, one may assume
that regulating the spacial area on the boundary set amounts to regulating
the super Yang-Mills theory on $M'\sim \bar{\cal G}\cap D_{\Gamma}$ with 
an UV cutoff $\epsilon$. This assumption can be verified directly by 
comparing the roles of the regulator $\epsilon$ in both the boundary- and 
bulk- theories, as described in section 2 of ref.\cite{SW}. For an independent 
check, one may also determine whether such an assumption will lead to the 
correct information density, which can be done following the discussion in 
section 3 of \cite{SW}. Thus, let us think of the regulated boundary theory 
as a theory living on a lattice whose cells have the size $\epsilon$. The 
total number of cells in the spacial directions can be estimated roughly by 
$(\Delta U)^{3}/\epsilon^{3}$, so for a large $N$ gauge theory on the lattice, 
the number of degrees of freedom $N_{dof}$ behaves as
\begin{equation}
N_{dof}\;\sim\;\frac{N^{2}(\Delta U)^{3}}{\epsilon^{3}}.
\label{dof}
\end{equation}
This together with (\ref{area}) as well as the relation between $R$ and 
$N$ given in (\ref{metric1}) leads to the desired result \cite{SW}
\begin{equation}
N_{dof}\;\sim\;A/G_{5}
\label{entropy}
\end{equation}
where $G_{5}\sim l_{s}^{8}g_{s}^{2}R^{-5}$ is the 5-dimensional Newton's
constant. Since each degree of freedom stores a single bit of information, 
the information density is given by $N_{dof}/A\sim 1/G_{5}$, and therefore
it cannot exceed one bit per Planck area. Note that as $\epsilon\rightarrow
0$, both the area (\ref{area}) and the number of degrees of freedom
(\ref{dof}) will blow up. However, even when such a cutoff is removed, the 
information density remains finite, bounded by the inverse Planck
area $1/G_{5}$.

Let us end with a brief discussion on the extreme case mentioned in section
3, where each component of $D_{\Gamma}={\bf S}^{4}-\Lambda_{\Gamma}$ collapses. 
Suppose that we have $p$ components of the boundary set $\bar{\cal G}\cap 
D_{\Gamma}$ and when $D_{\Gamma}$ collapses, this boundary set becomes $p$
cusps living in $\Lambda_{\Gamma}$. In that case, the length 
scale $\Delta U$ in (\ref{area})--(\ref{dof}) no longer keeps finite when the 
cutoff is removed, since as $\epsilon\rightarrow 0$, the Euclidean volume of 
the regulated boundary $\bar{\cal G}\cap T_{\epsilon}$ 
\begin{equation}
\int_{\bar{\cal G}\cap T_{\epsilon}} dT\,dX_{1}\,dX_{2}\,dX_{3}\;\approx\;
(\Delta U)^{4}
\label{e-vol}
\end{equation}
will be contributed from $p$ connected parts, each of which behaves as 
$\epsilon^{4}$ and thus $\Delta U\sim p^{1/4}\epsilon\rightarrow 0$. It
follows from this and (\ref{area})--(\ref{dof}) that 
\begin{equation}
A\;\sim\;p^{3/4}R^{3},\;\;\;\;\;\; Vol(X_{\epsilon})\;\sim\; p R^{5},
\;\;\;\;\;\; N_{dof}\;\sim\;p^{3/4}N^{2}, 
\label{collapse}
\end{equation}
so that removing the cutoff $\epsilon$ will not render the regulated 
quantities $A$, $Vol(X_{\epsilon})$ as well as $N_{dof}$ divergent. This 
confirms our observation made in section 3 that the extreme case itself
can be viewed as a regulated theory, in which no further regulating
procedure (e.g. introducing a cutoff $\epsilon$ as above) is needed. One 
may compare further $A\sim p^{3/4}R^{3}$ in (\ref{collapse}) with the 
usual expression for the regulated area \cite{SW} $A\sim R^{3}/\delta^{3}$, 
where $\delta$ is the dimensionless size of each cell on the discretized 
boundary. The comparison gives $\delta\sim p^{-1/4}$, which agrees with our 
earlier estimation (\ref{spacing}). The volume of $X$ estimated in 
(\ref{collapse}) is also consistent with the mathematical relation given in 
(\ref{vol-cusps}). The amount of information grows like $N_{dof}\propto
p^{3/4}$ when the number of cusps $p$ becomes large, showing that the cusps
are indeed capable of storing information. Finally, the factor $p^{3/4}$
in (\ref{collapse}) will be cancelled when computing the information density, 
which again leads to the expected result $N_{dof}/A\sim 1/G_{5}$. The 
considerations here may have a possible extension to the hybrid case in which 
only some of the components of $D_{\Gamma}$ collapse, and others do not.
\newpage

\end{document}